\documentclass[twocolumn]{aastex7}
\usepackage{newtxtext,newtxmath}
\usepackage{bm}
\usepackage{booktabs}

\begin{document}

\title{Induction Heating in Super-Earths: A Thermochemical Perspective}

\author[orcid=0000-0003-4404-8973]{Yihang Peng} 
\affiliation{Department of Geosciences, Princeton University, Princeton, NJ, USA}
\email[show]{yhpeng@princeton.edu (Y. P.)}

\author[orcid=0000-0003-4680-6774]{Kristina Kislyakova}
\affiliation{Department of Astrophysics, University of Vienna, Vienna, Austria}
\email{kristina.kislyakova@univie.ac.at}  

\author[orcid=0009-0003-0497-4961]{Donghao Zheng}
\affiliation{Department of Geosciences, Princeton University, Princeton, NJ, USA}
\email{donghao.zheng@princeton.edu}  

\author[orcid=0000-0002-3131-6169]{Zhongtian Zhang}
\affiliation{Department of Geosciences, Princeton University, Princeton, NJ, USA}
\affiliation{Earth and Planets Laboratory, Carnegie Institution for Science, Washington, DC, USA}
\email{zzhang10@carnegiescience.edu}

\author[orcid=0000-0001-5441-2797]{Jie Deng}
\affiliation{Department of Geosciences, Princeton University, Princeton, NJ, USA}
\email[show]{jie.deng@princeton.edu (J. D.)}  

\begin{abstract}

Electromagnetic induction heating has recently been proposed as an important internal heat source in the mantles of rocky exoplanets. However, its dependence on planetary interior properties remains poorly constrained. Here we construct electrical conductivity profiles for super-Earth mantles considering different temperatures and compositions, and evaluate induction heating in super-Earth mantles in both solid and partially molten states. We find that high mantle temperature, iron content, and melt fraction all suppress the overall induction heating efficiency due to increased mantle conductivity and magnetic shielding. In GJ~486b, induction heating likely exceeds both radiogenic heating and tidal heating, driving persistent surface volcanism and early volatile depletion, whereas HD~3167b and GJ~357b experience insignificant induction heating due to weak stellar magnetic fields. Our findings highlight induction heating as a critical factor in the thermal and atmospheric evolution of close-in super-Earths around magnetically active stars.

\end{abstract}

\keywords{\uat{Super-Earths}{1655} --- \uat{Planetary interior}{1248} --- \uat{Star-planet interactions}{2177} --- \uat{Exoplanet dynamics}{490} --- \uat{Magnetic fields}{994}}


\section{Introduction}

Among the over 5,800 confirmed exoplanets, 1--10 $M_\oplus$ rocky planets (super-Earths) are common, potentially habitable, and often Earth-like in structure \citep{valencia_inevitability_2007, valencia_radius_2007,  bloh_habitability_2007, seager_exoplanet_2013}. Heat production within super-Earth mantles directly governs mantle degassing through volcanism, which facilitates atmosphere formation
 and volatile cycling essential for habitability \citep{godolt_habitability_2019, unterborn_mantle_2022, luo_radiogenic_2024}. In addition to radiogenic and tidal heating, induction heating occurs when a planet interacts with a time-varying stellar magnetic field, inducing eddy currents and electromagnetic heat production in its mantle \citep{kislyakova_magma_2017} or atmosphere \citep{strugarek_ohmic_2025}. Under some conditions, planets can experience substantial induction heating comparable to radiogenic heating and tidal friction \citep{kislyakova_magma_2017, kislyakova_effective_2018, noack_interior_2021}, potentially driving mantle melting and enhanced volcanic outgassing in both super-Earths \citep{guenther_searching_2020, kislyakova_electromagnetic_2020} and smaller rocky planets \citep{kislyakova_magma_2017, kislyakova_effective_2018, kislyakova_induction_2023}.

While astronomical parameters controlling induction heating have been well studied \citep[e.g., stellar magnetic field, planetary orbital period, and orbital inclination,][]{kislyakova_effective_2018, kislyakova_electromagnetic_2020, kislyakova_induction_2023}, the role of planetary interior properties remains largely unexplored. The thermochemical properties of the mantle materials may influence induction heating by varying the electrical conductivity. The conductivity controls the skin depth $\delta = \sqrt{\frac{2}{\sigma \omega \mu}}$ (where $\sigma$ is conductivity, $\mu$ is the magnetic permeability, and $\omega$ is the frequency of the magnetic field), which determines the penetration depth of the external magnetic field and the spatial distribution of induction heating \citep{davies_conduction_1990, kislyakova_magma_2017}. Most previous studies have adopted Earth-like conductivity models for induction heating simulations \citep[e.g.,][]{kislyakova_magma_2017, kislyakova_electromagnetic_2020}. \citet{noack_interior_2021} considered variations in planetary mass and bulk iron fraction, as well as the resulting variations in mantle temperature and conductivity with mantle iron content assumed to be Earth-like. On the one hand, exoplanetary mantles may exhibit a wide range of chemical compositions, including iron contents from approximately 0 to 25 wt\% \citep{hatalova_compositional_2025}, which are expected to influence mantle conductivity \citep{Yoshino2010a, yoshino_effect_2012}. A comparable diversity is observed among terrestrial planets in the Solar System \citep[see Fig.~1 in][]{righter_terrestrial_2011}. On the other hand, while a surface temperature equal to the equilibrium temperature is usually assumed in planetary interior models \citep[e.g.,][]{kislyakova_electromagnetic_2020}, planetary evolution is likely to involve cooling from an initially molten state caused by accretionary heating, and mantle potential temperature can vary significantly even if the surface temperature is fixed at that given by the radiative equilibrium \citep[e.g.,][]{stixrude_melting_2014, vazan_contribution_2018, sahu_unveiling_2025}. Here, potential temperature refers to the temperature that a mantle parcel would have if adiabatically brought to surface, and it serves as a convenient anchor point for specifying physical thermal states of planetary mantles. The effect of mantle iron content and potential temperature on induction heating has not been systematically investigated in previous studies.

Furthermore, sufficiently strong induction heating may lead to partial melting in planetary mantles and even the formation of magma oceans (MOs) \citep{kislyakova_magma_2017, kislyakova_effective_2018}. Due to the significant contrast in electrical conductivity between rocks and silicate melts \citep{zhang_electrical_2021, Yoshino2010a}, the presence of partial melt and melt fraction may substantially influence induction heating patterns. Although previous studies have calculated the induction heating efficiency within fully molten mantles with rough conductivity profiles \citep{kislyakova_magma_2017, kislyakova_effective_2018}, a more realistic model is required to evaluate the response of induction heating to the onset of partial melting and the presence of MOs.

In this study, we developed new electrical conductivity profiles for super-Earth mantles across mantle potential temperature, iron content, and melt fraction, based on extensive experimental data and a recently updated model of planetary interior structures \citep{zheng_cation_2025}. We present a comprehensive examination of the effect of above thermochemical properties on the efficiency of induction heating in super-Earths.

\section{Methods Summary}

We investigate induction heating in super-Earths by combining planetary interior modeling \citep{zheng_cation_2025}, laboratory-based electrical conductivity models, and a refined electromagnetic induction framework. Electrical conductivities are calculated for the major mantle minerals (olivine, wadsleyite, ringwoodite, bridgmanite, and post-perovskite), incorporating both ionic and small-polaron conduction mechanisms with explicit dependence on pressure, temperature, and iron content \citep{yoshino_effect_2012, yoshino_electrical_2013}. The effect of partial melting on conductivity is included using experimental constraints on basaltic melts \citep{tyburczy_electrical_1983} and simple mixing relations \citep{waff_theoretical_1974}. Induction heating is then evaluated by solving the electromagnetic response of a spherically symmetric planet to a time-varying stellar magnetic field \citep{parkinson_introduction_1983, kislyakova_magma_2017}. Further methodological details are provided in Appendix~\ref{sec:method}.

\section{Results and Discussion}\label{sec:RD}

\subsection{Conductivity profiles of super-Earth mantles}\label{sec:sigma}

\begin{figure*}[ht!]
\centering
\includegraphics[width=0.8\textwidth]{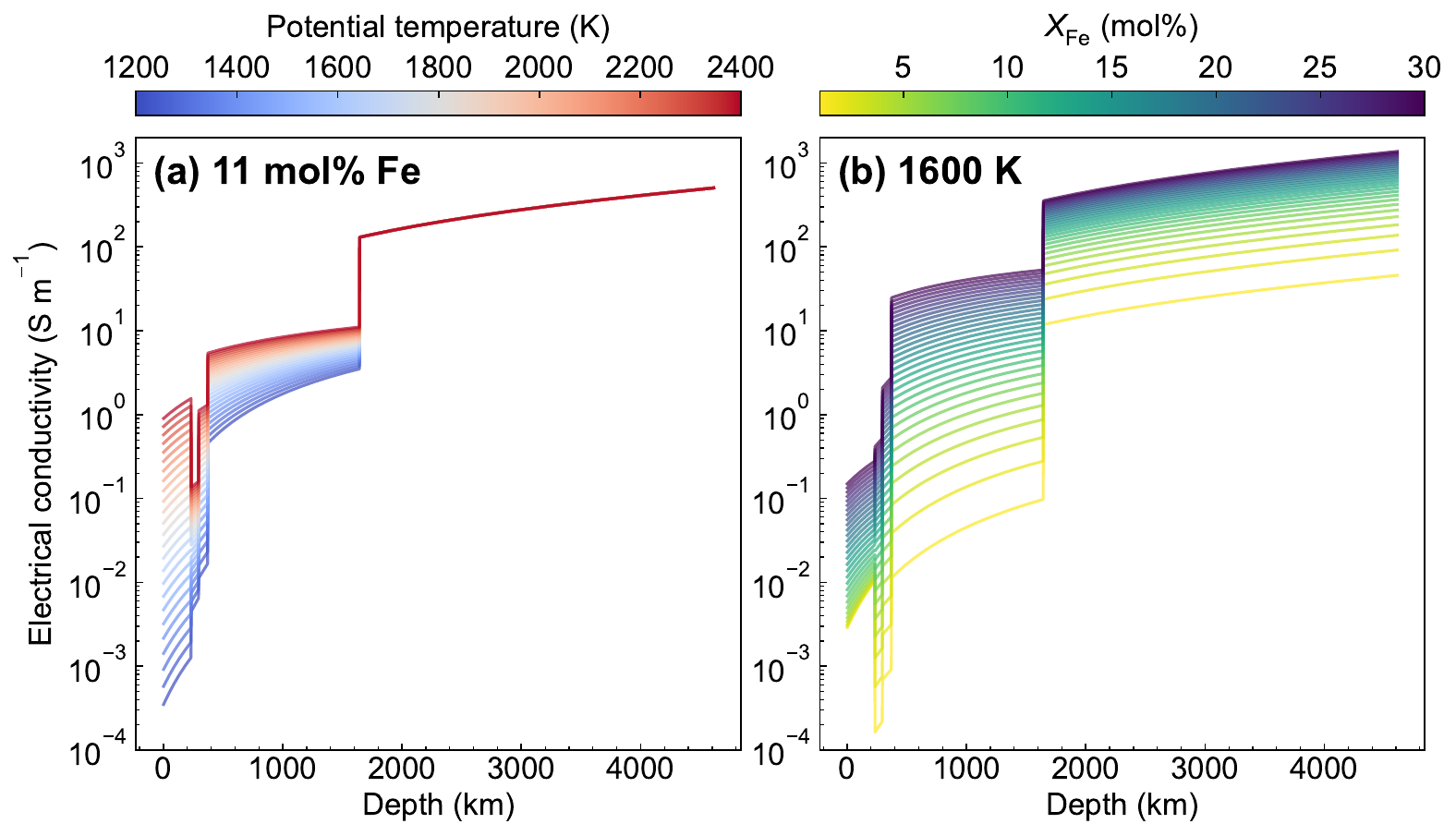}
\caption{Mantle conductivity profiles for a 4 $M_\oplus$ super-Earth with an Earth-like core mass fraction (0.32). (a) Conductivity profiles with various potential temperatures. (b) Conductivity profiles with various Fe fractions ($X_\mathrm{Fe}$) in mantle silicates.}\label{fig:sigma}
\end{figure*}

We modeled mantle conductivity profiles for a 4 $M_\oplus$ super-Earth with an Earth-like core mass fraction (0.32) at varying potential temperature ($T_p$) (Figure~\ref{fig:sigma}a) and iron content ($X_{\rm Fe}$) (Figure~\ref{fig:sigma}b). Conductivity generally increases with depth, with jumps at phase transition boundaries of silicate minerals. Different mineral phases show significantly different temperature dependencies of conductivity (Figure~\ref{fig:sigma}a), which is quantified by activation energy and related to the conduction mechanisms. In a dry and iron-bearing mantle, wadsleyite, ringwoodite, and bridgmanite are primarily governed by small polaron conduction, involving thermally activated hopping of charged polarons between Fe\textsuperscript{2+} and Fe\textsuperscript{3+} sites \citep{yoshino_electrical_2013, Yoshino2016a}. The activation energy of this mechanism is sensitive to ferric iron content. In bridgmanite, which preferentially incorporates Fe\textsuperscript{3+} \citep{dorfman_effects_2020}, small polaron conduction has a low activation energy of $\sim$0.5 eV \citep{Xu1998a, xu_laboratory-based_2000, Yoshino2016a}. In contrast, olivine typically hosts much less Fe\textsuperscript{3+} \citep{canil_distribution_1996} and exhibits a higher energy barrier for small polaron conduction (1--2 eV) \citep{yoshino_effect_2012}. Additionally, ionic conduction with a very high activation enthalpy (2.31 eV) becomes the dominated conduction mechanism in olivine under high temperatures \citep{yoshino_effect_2009}. Therefore, the conductivity in the upper mantle increases by nearly three orders of magnitude as $T_p$ increases from 1200 K to 2200 K, whereas the lower-mantle conductivity varies by less than one order of magnitude over the same $T_p$ range. $X_{\rm Fe}$ has a strong positive effect on conductivity, which is broadly consistent across different mineral phases (Figure~\ref{fig:sigma}b). This can be attributed to the reduced average Fe\textsuperscript{3+}-Fe\textsuperscript{2+} distance, which lowers the energy barrier for small polaron hopping \citep{yoshino_effect_2012, ohta_electrical_2008}. The conductivity of post-perovskite has negligible impact on the induction heating due to the small skin depth of external magnetic fields (Appendix~\ref{sec:ppv}).

\subsection{Induction heating in super-Earth mantles}\label{sec:heating}

\begin{table*}
\caption{Physical parameters of exoplanets GJ~486b, HD~3167b, GJ~357b and their host stars adopted from \citet{trifonov_nearby_2021}, \citet{gandolfi_transiting_2017}, \citet{kislyakova_electromagnetic_2020}, and \citet{luque_planetary_2019}.}\label{tab:parameters}
\centering
\begin{tabular}{ccccc}
\toprule
\textbf{Parameter} & \textbf{GJ~486b} & \textbf{HD~3167b} & \textbf{GJ~357b} & \textbf{Unit} \\
\midrule
Stellar mass & $0.323 \pm 0.015$& $0.877 \pm 0.024$ & $0.342 \pm 0.011$ & $M_\odot$ \\
Stellar radius & $0.328 \pm 0.011$& $0.835 \pm 0.026$ & $0.337 \pm 0.015$ & $R_\odot$ \\
Stellar equilibrium temperature & $3340 \pm 54$& $5286 \pm 40$ & $3505 \pm 51$ & K \\
Stellar rotation period & $130.1 \pm 1.4$& $23.52 \pm 2.87$ & $78 \pm 2$ & days \\
Planetary mass & $2.82 \pm 0.11$& $5.69 \pm 0.44$ & $1.84 \pm 0.31$ & $M_\oplus$ \\
Planetary radius & $1.305 \pm 0.065$& $1.574 \pm 0.054$ & $1.217 \pm 0.084$ & $R_\oplus$ \\
Planetary equilibrium temperature & $701 \pm 13$& $1759 \pm 20$ & $525 \pm 10$ & K \\
Semi-major axis & $3.727 \pm 0.056$& $3.766 \pm 0.135$ & $7.52 \pm 0.43$ & $R_\odot$ \\
Orbital inclination & 88.4& 85 & 88.496 & degree \\
\bottomrule
\end{tabular}
\end{table*}

We systematically examine the effect of $T_p$, $X_\mathrm{Fe}$, and partial melting on the induction heating in three representative super-Earths: GJ~486b, HD~3167b, and GJ~357b (Table~\ref{tab:parameters}).

\subsubsection{GJ~486b}\label{sec:GJ486b}

\begin{figure*}[ht!]
\centering
\includegraphics[width=0.8\textwidth]{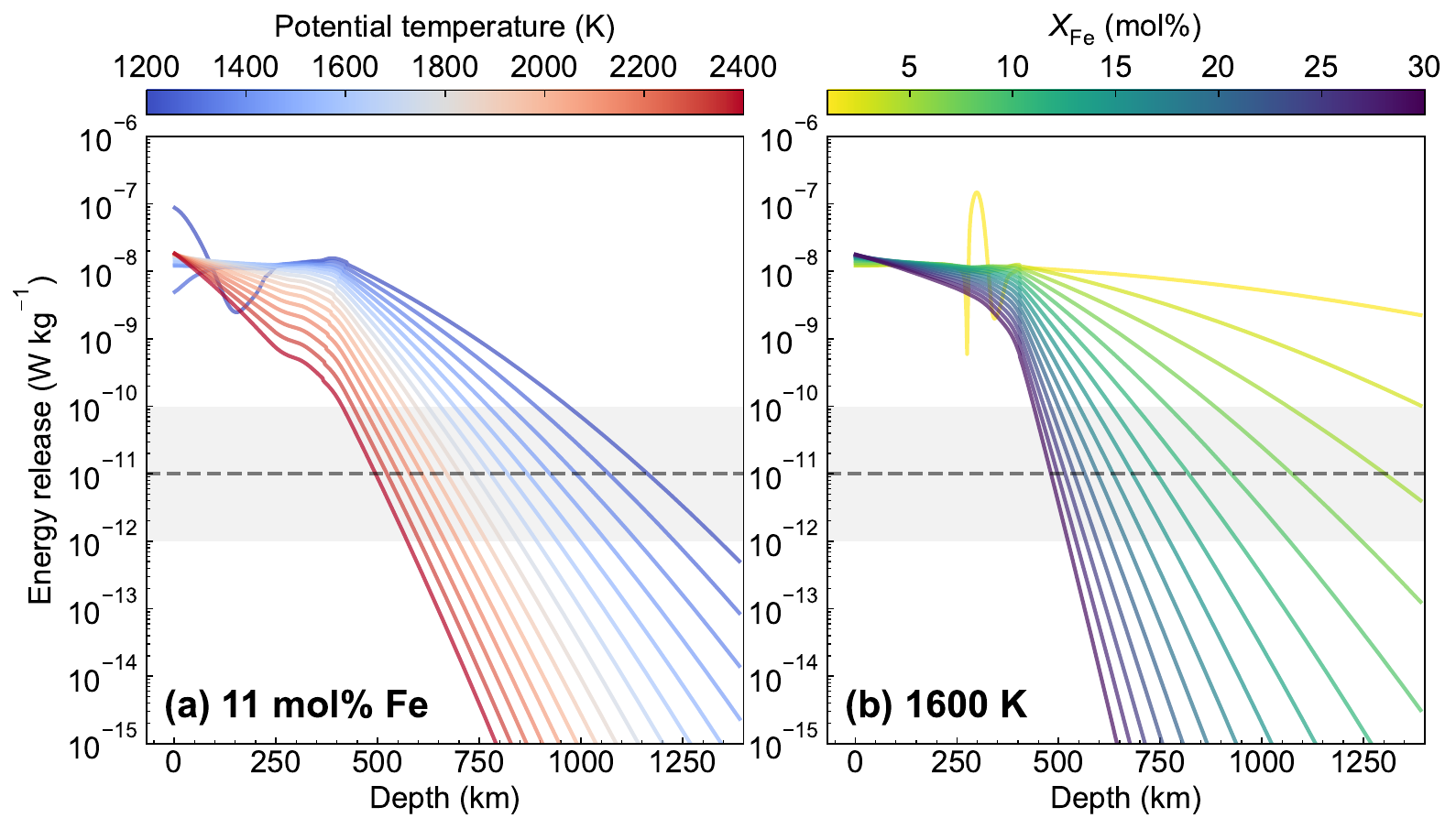}
\caption{Energy release per unit mass due to induction heating as a function of depth inside GJ~486b considering different potential temperatures (a) and Fe fraction ($X_\mathrm{Fe}$) in mantle silicates (b). The critical value of energy release that can keep the mantle rock molten ($10^{-11}$ W\;kg$^{-1}$) and its plausible range \citep{kislyakova_effective_2018, guenther_searching_2020} are shown as dashed lines and shaded regions, respectively.}\label{fig:heating}
\end{figure*}

With a mass of $\sim$2.8~$M_\oplus$, GJ~486b is a recently discovered super-Earth orbiting an M3.5 dwarf star \citep{trifonov_nearby_2021}, which has received significant attention regarding its atmospheric composition, interior structure, and thermal evolution \citep{weiner_mansfield_no_2024, sahu_unveiling_2025}. The host star, GJ~486, is a fully convective star with a mass less than 0.35~$M_\odot$ and possesses a strong magnetic field of $1.6 \pm 0.3$~kG \citep{moutou_spirou_2017}. The induction heating can be highly efficient around such strongly magnetized stars, potentially driving extreme surface volcanism and the formation of MOs \citep{kislyakova_effective_2018}. Moreover, recent radio observations of GJ~486 show very low magnetic obliquity and stellar inclination \citep{pena-monino_searching_2025}, suggesting that the orbit of GJ~486b is significantly inclined with respect to both the stellar rotation axis and the magnetic dipole axis, a configuration that can maximize the efficiency of induction heating \citep{kislyakova_effective_2018}. Given its potentially strong induction heating and the well-constrained astronomical parameters, GJ~486b represents an ideal target for exploring the interior effects on induction heating efficiency.

Figure~\ref{fig:heating} presents the calculated induction heating power per unit mass as a function of the depth inside GJ~486b under varying $T_p$ and $X_{\mathrm{Fe}}$. Induction heating exhibits a surface-confined peak whose magnitude remains nearly constant across a wide range of parameter variations, whereas the heating rate generally decreases with increasing depth. As $T_p$ and $X_{\mathrm{Fe}}$ decrease, the overall efficiency of induction heating increases systematically. These features are primarily governed by the interior profiles of electrical conductivity, magnetic field strength, and current density obtained from our model, which are discussed in detail in Appendix \ref{sec:app-heating}.

\begin{figure*}[ht!]
\centering
\includegraphics[width=\textwidth]{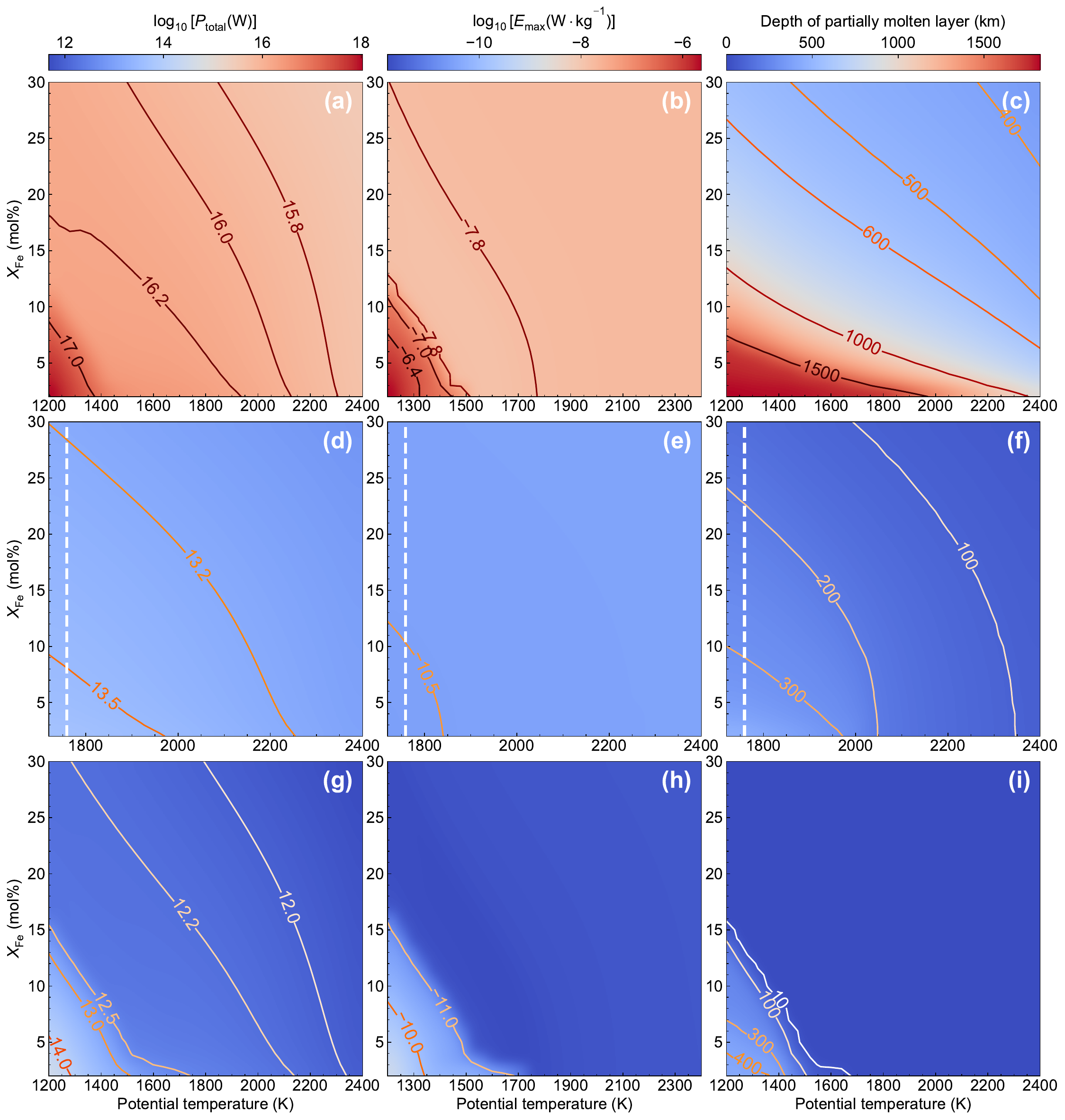}
\caption{Color maps and contour plots of three metrics of induction heating efficiency as a function of potential temperature and mantle Fe contents ($X_\mathrm{Fe}$) in GJ~486b (a--c), HD~3167b (d--f), and GJ~357b (g--i). (a, d, g) Total induction heating power inside the planet; (b, e, h) Maximum depth-dependent energy release per unit mass; (c, f, i) Depth of the partially molten mantle layer that has an energy release exceeding the critical value of $10^{-11}$ W\;kg$^{-1}$. The white dashed lines mark the equilibrium temperature of HD~3167b \citep[1759K, ][]{kislyakova_electromagnetic_2020}.}\label{fig:contour}
\end{figure*}

This combined effect of $T_p$ and $X_\mathrm{Fe}$ on total induction heating power in GJ~486b is shown in Figure~\ref{fig:contour}a, where values increase from $4.0 \times 10^{15}$~W at low $T_p$ and $X_\mathrm{Fe}$ to $8.5 \times 10^{17}$~W at high $T_p$ and $X_\mathrm{Fe}$. For comparison, the present-day radiogenic heat production in the Earth's mantle is $\sim 2.1 \times 10^{13}$~W, 2--4 orders of magnitude lower than the induction heating in GJ~486b. The surface heat flux generated by induction heating (5--980~W\;m$^{-2}$) also significantly exceeds that of Jupiter's satellite Io (2.24~W\;m$^{-2}$), where tidal dissipation drives the most intense volcanic activity in the Solar System \citep{lainey_strong_2009}. Therefore, induction heating may strongly dominate the internal heat budget of GJ~486b, far surpassing all other potential heat sources.

We consider a critical threshold of $10^{-11}$~W\;kg$^{-1}$, estimated as the heating rate required for mantle material to accumulate enough energy over geological timescales to reach the local solidus, to assess the possibility of mantle partial melting due to induction heating. This value depends on multiple variables and should be regarded only as an rough estimate, thus we consider a plausible range of 10$^{-12}$--10$^{-10}$ W\;kg$^{-1}$ (Figure~\ref{fig:heating}) following previous studies \citep{kislyakova_magma_2017, kislyakova_electromagnetic_2020, noack_interior_2021}. As shown in both Figure~\ref{fig:heating} and Figure~\ref{fig:contour}b, the maximum energy release by induction heating near planetary surface exceeds this threshold by at least three orders of magnitude over the entire parameter space. The depth over which this threshold is surpassed, i.e., the depth of possibly partially molten mantle, can even exceed 1500 km under cool and iron-poor conditions (Figure~\ref{fig:contour}c). It is very likely that induction heating has generated and sustained partial melting in the mantle of GJ~486b and even large-scale MOs, which will be further examined in Section~\ref{sec:molten}.

\subsubsection{HD~3167b}\label{sec:HD3167b}

HD~3167b is a 5~$M_\oplus$ super-Earth orbiting a K0-type star. The stellar magnetic dipole strength has been estimated to be about 10 G or less \citep{kislyakova_electromagnetic_2020}. In a previous study, the induction heating in HD~3167b has already been calculated \citep{kislyakova_electromagnetic_2020}; however, we recently noticed that previous models have some numerical issues and the actual converged heating power is around one order of magnitude lower than previously reported (Figure~\ref{fig:correct}).

We therefore re-evaluated the induction heating efficiency in HD~3167b using a revised numerical scheme and updated conductivity profiles that account for a broad range of mantle $X_\mathrm{Fe}$ and $T_p$. Due to the extremely high equilibrium temperature of 1759 K \citep{gandolfi_transiting_2017}, we adjusted the range of $T_p$ accordingly during the calculations. The resulting total induction heating power and the maximum energy release per unit mass are shown in Figure~\ref{fig:contour}d,e,f. Consistent with the case of GJ~486b, increasing the mantle temperature and iron content both reduce the efficiency of induction heating by enhancing conductivity. Across the entire parameter space explored, the total induction heating power in HD~3167b ranges from $6.3 \times 10^{12}$ W to $5.0 \times 10^{13}$ W (Figure~\ref{fig:contour}d), which is significantly lower than the previously reported value of $1.8 \times 10^{14}$ W  \citep{kislyakova_electromagnetic_2020}. Although a 10 G stellar magnetic field can still drive partial melting within a shallow layer (30--400 km) beneath the surface (Figure~\ref{fig:contour}f), reducing the magnetic field to a moderate value of 5 G eliminates any layers where the energy release surpasses the critical threshold (Figure~\ref{fig:5G}). Therefore, the role of induction heating in promoting partial melting and mantle outgassing of HD~3167b is much weaker than previously estimated. Furthermore, given the high surface temperature, the temperature of the upper mantle may already approach or exceed the mantle solidus (Figure~\ref{fig:solidus}), making it particularly important to assess the additional impact of partial melting on induction heating (Section~\ref{sec:molten}).

\subsubsection{GJ~357b}\label{sec:GJ357b}

GJ~357b is an Earth-size planet with a radius of $\sim$1.2~$R_\oplus$ and a mass of $\sim$1.8~$M_\oplus$, orbiting a nearby M2.5 dwarf \citep{luque_planetary_2019}. Due to its long rotation period ($78 \pm 2$ days) and extremely low levels of X-ray activity \citep{modirrousta-galian_gj_2020, luque_planetary_2019}, GJ~357 may have a weak magnetic dipole field of less than 100~G. In addition, due to the larger orbital semi-major axis and the more rapid spatial decay of the stellar magnetic field (attributable to the smaller size of its host star), the field strength at GJ~357b's orbit is only 0.26~G (for a stellar dipole of 100~G)---about half that at HD~3167b's orbit (0.52~G for a 10~G dipole) and merely 2\% of that at GJ~486b's orbit (14.4~G for a 1600~G dipole).

Therefore, our calculations indicate that the induction heating power inside GJ~357b is generally lower than that of HD~3167b, remaining below $10^{13}$~W except in scenarios where the mantle is both extremely cool and iron-poor (Figure~\ref{fig:contour}g). Furthermore, the actual stellar magnetic field is likely to be weaker than the assumed upper bound 100~G, and the orbital inclination with respect to the stellar dipole axis remains unknown. In the calculations, we adopt similar inclination values as for GJ~486b and HD~3167b (Table~\ref{tab:parameters}) to estimate an upper limit of induction heating efficiency. Taking these factors into account, we conclude that induction heating is unlikely to significantly contribute to the internal heat budget or trigger volcanic activity on GJ~357b or similar exoplanets. This outcome highlights a compensating effect: while smaller and cooler late-type M dwarfs may possess stronger magnetic fields, their magnetic fields also decay more rapidly with distance compared to those of Sun-like stars (e.g., HD~3167). As a result, efficient induction heating within the planets orbiting around M dwarfs often requires very close-in orbits---as seen in the case of GJ~486b, which has a semi-major axis of only 0.017~AU.

\subsection{Response of induction heating to mantle partial melting}\label{sec:molten}

\begin{figure*}[ht!]
\centering
\includegraphics[width=0.8\textwidth]{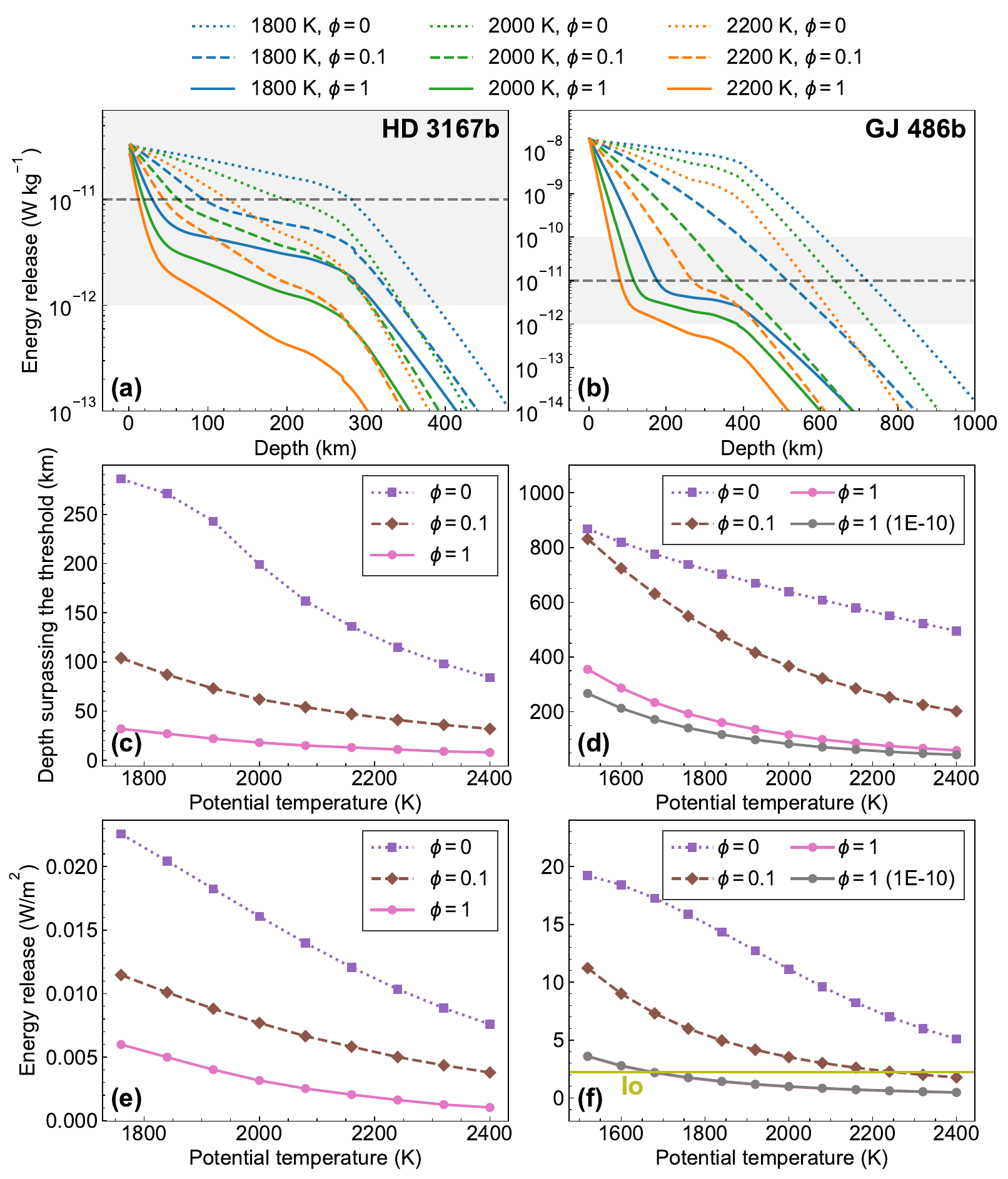}
\caption{(a,b) Energy release due to induction heating as a function of depth in HD~3167b (a) and GJ~486b (b) with the presence of mantle partial melting with different potential temperatures. (c, d) The depth of the partial melt that can be sustained by induction heating in HD~3167b (c) and GJ~486b (d). (e, f) the surface heat flux produced by induction heating as a function of potential temperature in HD~3167b (e) and GJ~486b (f). The energy release due to tidal heating in Io \citep{lainey_strong_2009} is shown for comparison in (f). The gray curves labeled 1E-10 adopt a high critical threshold of $10^{-10}$~W\;kg$^{-1}$ for mantle partial melting. All other calculations use a threshold value of $10^{-11}$~W\;kg$^{-1}$. The pink and gray curves in panel f almost overlap. The mantle iron content is set to be an Earth-like value of 11 mol\%.}\label{fig:melting}
\end{figure*}

The strong induction heating in GJ~486b as well as the high equilibrium temperature of HD~3167b suggests that extensive partial melts may occur in the upper mantles of both planets. Due to the sharp melt-rock conductivity contrast, it is essential to investigate how induction heating responds to the presence of partial melting and MOs.

We computed the induction heating of GJ~486b and HD~3167b using different melt fractions ($\phi = 0$, 0.1, and 1) and an Earth-like mantle iron content of 11~mol\%. We iteratively adjusted the melt layer depth based on the heating profile until convergence (Appendix~\ref{sec:iterative}). Due to the significantly higher conductivity of melt compared to minerals, partially molten layers exhibit a much stronger shielding effect against high-frequency external magnetic fields (Figure~\ref{fig:melting}a,b). This leads to a reduced skin depth and a much faster decay of the magnetic field within the molten region, as compared to the solid mantle. As a result, the energy released by induction heating decreases sharply with depth in the presence of melt, ultimately reducing both the total induction heating power and the thickness of the affected region (Figure~\ref{fig:melting}).

\subsubsection{HD~3167b}

In the case of HD~3167b, even a melt fraction of only 0.1 is sufficient to confine the partially molten mantle to a shallow crust zone only a few tens of kilometers thick (Figure~\ref{fig:melting}c). In addition, the surface temperature of HD~3167b ($\sim$1759 K) may be high enough to support a surface MO itself (Figure~\ref{fig:solidus}), which would cause strong shielding effect on its induction heating. Such limited heating is unlikely to have a significant effect on mantle degassing processes. This result is consistent with current observational data, which do not support ongoing volcanic activity on HD~3167b \citep{guenther_searching_2020}. Instead, the presence of a mineral atmosphere is possible due to surface rock melting caused by the stellar radiation \citep{guenther_searching_2020, miguel_compositions_2011} as well as the surface induction heating. Our findings suggest that for Sun-like stars with magnetic field strengths on the range of 1--10 G, induction heating is unlikely to significantly influence the planetary heat budget or drive volcanic activity on rocky planets.

\subsubsection{GJ~486b}\label{sec:melt486}

For GJ~486b, owing to the stronger external magnetic field, a partially molten layer is not sufficient to effectively shield the field, especially at lower $T_p$ where the conductivity of the melt is reduced. Even in the case of a fully molten MO, induction heating can sustain melting to depths of up to 300~km when $T_p$ is Earth-like at 1600~K---which is also the present-day potential temperature predicted by previous thermal evolution models for GJ~486b \citep{sahu_unveiling_2025}. Under this condition, induction heating can produce an additional surface heat flux of at least 3~W\;m$^{-2}$ (depending on the exact melt fraction, Figure~\ref{fig:melting}f). This value surpasses the surface heat flux of Io, which indicates intense volcanic activity and early mantle degassing in GJ~486b.

We now assess whether induction heating can lead to the formation of a global MO in the upper mantle of GJ~486b. The degree of melting depends on the balance between internal heat generation and surface heat loss. Once partial melting occurs, the buoyant melt \citep{ohtani_melting_1995} can percolate upward and efficiently transport heat to the surface where it crystallizes and releases latent heat. Previous studies have shown that, in a Mars-sized planetary embryo, percolation-induced heat flux at a melt fraction of 0.3 can be over 500~W\;m$^{-2}$ \citep{zhang_two-phase_2021}; and even with heat generation of $\sim$300~W\;m$^{-2}$, the body may remain largely solid (i.e., with melt fractions below 0.2) \citep{zhang_two-phase_2021}. Moreover, recent observations suggest that even under strong tidal heating exceeding 2~W\;m$^{-2}$ \citep{lainey_strong_2009}, Io does not host a shallow global MO and retains a mostly solid mantle \citep{park_ios_2024}. Considering the comparable magnitude of its surface heat flux to Io and the stronger surface gravity of a super-Earth that facilitates melt migration, we infer that GJ~486b is also unlikely to host a global shallow MO.

However, we cannot completely rule out the existence of a subsurface MO in GJ~486b, for the following reasons: 1) The efficiency of melt-induced heat transport depends on several uncertain parameters, including melt viscosity and solid grain size \citep{zhang_two-phase_2021}; 2) Even in the presence of Rayleigh-Taylor instabilities, a dense crust overlying a molten mantle layer may persist over gigayear timescales \citep{miyazaki_stability_2024}, in which case heat is primarily released via volcanic eruptions (heat pipes) rather than percolation, leading to much lower heat loss efficiency \citep{miyazaki_subsurface_2022, miyazaki_stability_2024}; 3) Due to the limitations in melt conductivity models, our calculations in this section assumed an Earth-like mantle iron content. If GJ~486b has an iron-depleted mantle, the resulting decreased conductivity may enhance induction heating and help sustain the MO (Figure~\ref{fig:contour}a). It is worth noting that the potential MO may exhibit a non-adiabatic, locally inverted temperature profile: hot, buoyant layers may reside above cooler, denser ones due to the steep decay of induction heating with depth, representing a gravitationally stable stratification and suggesting that convection within the MO is likely to be limited. This unique structure may to some extent hinder the exchange of material and energy between the planetary interior and surface, potentially impacting the elemental cycling, mantle dynamics, and thermal evolution of GJ~486b. Such non-adiabatic temperature profiles are also predicted in our simple self-consistent calculations (Appendix \ref{sec:tempsc}), which also demonstrate that even when the deviation from an adiabatic temperature profile caused by induction heating is included, the heating results remain close to the original values, and therefore the conclusions of this study remain solid.

\subsection{Implications for the atmosphere of GJ~486b}

Intense induction heating would trigger strong volcanic activity on GJ~486b and cause substantial volatile release from the hundreds-of-kilometers-deep mantle region affected by the heating, which may lead to an earlier formation of a thick atmosphere \citep{kislyakova_magma_2017, kislyakova_effective_2018, noack_interior_2021}. However, additional electromagnetic interaction between a close-in planet and its host star may occur once the atmosphere formed. The strong X-ray and ultraviolet (XUV) radiation can generate a conductive planetary ionosphere with high Pedersen conductivity \citep{strugarek_ohmic_2025}. As a result, the upper atmosphere exposed to a time-varying stellar magnetic field may undergo substantial Ohmic heating, even exceeding the heat production by stellar XUV \citep{strugarek_ohmic_2025}.

The intrinsic magnetic field of the planet strongly influences the Pedersen conductivity and thus the induction heating in the upper atmosphere \citep{strugarek_ohmic_2025}. \citet{luo_radiogenic_2024} suggest that in super-Earths with high core-mantle differentiation pressures, traditionally lithophile heat-producing elements may become siderophile and preferentially partition into the core. This process may significantly enhance the heat flux across the core-mantle boundary and sustain strong magnetic dynamos. For a GJ~486b-sized super-Earth, the modeling results suggest a planetary magnetic field with a strength of 0.2--1.0~G to persist for over 10~Gyr \citep{luo_radiogenic_2024}. A magnetic field in this range can allow stellar magnetic fields to penetrate below the atmospheric sonic point and cause intense atmospheric Ohmic heating \citep{strugarek_ohmic_2025}, accelerating the thermal escape of the atmosphere. As the atmosphere becoming thinner, stronger stellar magnetic fields may again penetrate into the planetary interior, triggering the outgassing of remaining mantle volatiles. The replenishment of the atmosphere can further amplify both atmospheric Ohmic heating and thus the efficiency of atmospheric depletion.

Given the 6.6 Gyr history of the GJ~486 system \citep{diamond-lowe_high-energy_2024} and the long-term atmospheric escape processes, it is highly probable that GJ~486b has lost its atmosphere and become an airless planet. This interpretation is remarkably consistent with the latest observations by \citet{weiner_mansfield_no_2024}. JWST eclipse observations (5--12~$\mu$m) of GJ~486b report a dayside temperature of $865 \pm 14$ K, nearly the bare-rock maximum ($0.97 \pm 0.01$). The spectrum matches either an airless basaltic/metal-rich surface or a tenuous atmosphere ($<$1\% H$_2$O, $<$1~ppm CO$_2$) \citep{weiner_mansfield_no_2024}, and the near-maximum temperature points to a hot subsurface or even a global shallow MO. We propose that rapid atmosphere formation and escape driven by the interplay of the mantle and atmospheric induction heating is likely a widespread process among close-in super-Earths orbiting strongly magnetized stars.

\section{Conclusions}\label{sec:conclusions}
In this study, we systematically explored how variations in mantle temperature, iron content, and partial melting affect induction heating in super-Earth planets. Our results reveal that the efficiency and spatial distribution of induction heating are highly sensitive to the planetary thermochemical structure. Heating consistently peaks at the planetary surface and can be well explained by the interior profiles of electrical conductivity, magnetic field, and current density. In GJ~486b with the assumption of a fully solid mantle, induction heating power ranges from $4.0 \times 10^{15}$~W to $8.5 \times 10^{17}$~W under plausible interior conditions. Induction heating in GJ~486b gradually weakens with increasing partial melt fraction in the mantle, but the resulting heat flux consistently exceeds the tidal dissipation on Io. On the contrary, the induction heating remains weak across the parameter space in HD~3167b and GJ~357b and probably other similar exoplanets under the influence of Sun-like stellar magnetic fields.

We find that strong volcanic activity and possibly a shallow MO can be sustained by induction heating alone for close-in super-Earths orbiting around strongly magnetized stars. We show that the combination of interior and atmospheric induction heating may promote early depletion of both interior volatiles and planetary atmosphere during long-term evolution, which align with recent observations and support the interpretation of GJ~486b as an airless planet. Overall, our study demonstrates that induction heating is an important heat source on rocky exoplanets orbiting magnetically active stars, strongly affecting their interior dynamics and atmospheric evolution. These findings highlight the need for future multi-disciplinary studies to fully understand the evolutionary consequences of induction heating.

\begin{acknowledgments}
We thank Sudeshna Boro Saikia and Lena Noack for helpful discussions. This work was funded by the National Science Foundation under Grant EAR-2242946 to J.D. The work reported in this paper was performed using the Princeton Research Computing resources at Princeton University which is consortium of groups led by the Princeton Institute for Computational Science and Engineering (PICSciE) and Office of Information Technology's Research Computing.
\end{acknowledgments}

\begin{contribution}

J.D. conceived the initial research concept and supervised the project. Y.P., K.K., and D.Z. carried out the modeling. Y.P. wrote the first draft. Y.P., K.K., Z.Z., D.Z., and J.D. all contributed to data analysis and writing the paper.

\end{contribution}

\appendix

\renewcommand{\thefigure}{A\arabic{figure}}
\setcounter{figure}{0}

\renewcommand{\thetable}{A\arabic{table}}
\setcounter{table}{0}

\section{Methods}\label{sec:method}

\subsection{Planetary interior model}

Three specific super-Earths (GJ~486b, HD~3167b, and GJ~357b) are selected for modeling in this study. The astrophysical parameters of these planets and their host stars are summarized in Table~\ref{tab:parameters}. The temperature, pressure, and density profiles of planetary mantles required for conductivity and induction heating calculations are obtained from our recently developed planetary interior model \citep{zheng_cation_2025}, using the planetary mass (Table~\ref{tab:parameters}), radius (Table~\ref{tab:parameters}), potential temperature, and core-mass fraction as input parameters. The core-mass fractions of GJ~486b, HD~3167b, and GJ~357b are assumed to be 0.448, 0.406, and 0.609 based on a previously reported mass-radius relation \citep{boujibar_super-earth_2020}.

\subsection{Electrical conductivity model}

Electrical conductivity of dry silicates in planetary mantles is contributed by two conduction mechanisms: ionic conduction ($\sigma_i$) and small polaron conduction ($\sigma_h$). The electrical conductivity as a function of pressure, temperature, and iron content can be calculated using the following Arrhenius model \citep{yoshino_effect_2012, yoshino_electrical_2013}:
\begin{equation}\label{eq:arrhen}
\sigma = \sigma_i + \sigma_h \notag = \sigma_{i,0} \exp \left(-\frac{\Delta H_i}{k_B T}\right) + \sigma_{h, 0} X_{\mathrm{Fe}} \notag\exp \left[ -\frac{\left(\Delta E_{h,0}-\alpha X_{\mathrm{Fe}}^{1 / 3}\right)+P\left(\Delta V_{h, 0}-\beta X_{\mathrm{Fe}}\right)}{k_B T}\right]
\end{equation}
where $k_B$ is the Boltzmann constant; $P$ is pressure; $T$ is temperature in K; $X_{\mathrm{Fe}}$ is the mole fraction of Fe in silicates defined as Fe/(Fe+Mg); $\sigma_{i,0}$ and $\sigma_{h,0}$ are the pre-exponential factors; $\Delta H_i$ is the activation energy of ionic conduction; $\Delta E_{h,0}$ and $\Delta V_{h,0}$ are the activation energy and activation volume of small polaron conduction observed at very low Fe concentrations; $\alpha$ and $\beta$ are constants \citep{yoshino_effect_2012}. Five major mineral phases are considered when calculating the conductivity profiles in a solid mantle: olivine, wadsleyite, ringwoodite, bridgmanite, and post-perovskite. The parameters for small polaron conduction in olivine, wadsleyite, and ringwoodite are reported by \citet{yoshino_effect_2012}. The parameters for ionic conduction in olivine are reported by \citet{yoshino_effect_2009}. For high pressure phases beyond olivine, the parameters for ionic conduction are largely undetermined \citep{yoshino_electrical_2013}, and the ionic conduction in the deep mantle is negligibly small \citep{Yoshino2016a}. Therefore, the ionic conduction is only considered in the olivine phase in our model. The conductivity model for bridgmanite is adopted from the experimental model \citep{Xu1998a} by adding the effect of Fe on the small polaron conduction \citep{yoshino_effect_2012}. The conductivity of post-perovskite is estimated upon the only available experimental results under $\sim$130 GPa \citep{ohta_electrical_2008} (Section~\ref{sec:sigma}). The conductivity of silicate melts is approximated using the experimental Arrhenius model for basaltic melts \citep{tyburczy_electrical_1983}. At low melt fractions (less than 15\%), the conductivity of a partially molten mantle with continuous grain boundary wetting can be estimated using the relation $\sigma \approx \frac{2}{3} \phi \sigma_m$ \citep{waff_theoretical_1974}, where $\phi$ is the melt fraction and $\sigma_m$ is the conductivity of the melt. This simple theoretical model agrees very well with the experimental results of the conductivity of melt-bearing peridotites \citep{yoshino_electrical_2010}. Compared to previous studies \citep[e.g.,][]{kislyakova_electromagnetic_2020}, we employed an updated conductivity model that includes more variable parameters such as pressure and iron content. However, when all parameters are held the same, the induction heating results are consistent with those from previous conductivity models (Figure~\ref{fig:correct}). For Fe-bearing silicates under the temperature condition of planetary interiors, magnetic ordering is absent and the materials are effectively paramagnetic \citep{aronson_magnetic_2007}. We therefore set the vacuum value of the magnetic permeability for all calculations.

\subsection{Induction heating model}\label{sec:induction}

We calculated the energy production of induction heating based on a numerical framework previously used by \citet{kislyakova_magma_2017, kislyakova_effective_2018} and \citet{kislyakova_electromagnetic_2020}. In this model, the planet is treated as a spherically symmetric body composed of concentric layers, each characterized by a uniform electrical conductivity. The electromagnetic response of the planet to a time-varying external magnetic field is computed by solving the induction equation in each layer, following the formalism of \citet{parkinson_introduction_1983}. This approach allows us to determine the magnetic field and the induced current within each layer. The volumetric rate of energy dissipation due to induction heating is then obtained from the local conductivity and current density. A detailed description of the mathematical formulation is provided in \citet{parkinson_introduction_1983}, with its application to exoplanets outlined in \citet{kislyakova_magma_2017}.

In this study, we improved the numerical accuracy and robustness  of the model by: 1) significantly increasing the resolution of the conductivity profiles and refining the layer discretization; 2) stabilizing the boundary condition of the reflection coefficient at the core-mantle boundary; 3) calculating the current density directly from a finite-difference curl of the magnetic field instead of the magnetic vector potential. We found that the previously reported results may contain numerical inaccuracies in both the total power of induction heating and its spatial distribution along the planetary radius (see Section~\ref{sec:HD3167b} and Figure~\ref{fig:correct}). Our updated approach ensures full numerical convergence and enables accurate quantification of induction heating across a broad range of planetary interior profiles.

\subsection{Self-consistent calculation of the partial melting depth}\label{sec:iterative}

We utilized an iterative calculation scheme to determine the final depth of the partially molten layer self-consistently. The initial depth of partial melting was set to 10~km. The corresponding conductivity profile was used to compute the induction heating profile, which was then compared to the critical threshold of $10^{-11}$~W\;kg$^{-1}$ to determine the updated depth of the molten layer. A more strict threshold of $10^{-10}$~W\;kg$^{-1}$ is also tested and does not affect our conclusions (Figure~\ref{fig:melting}d,f). This procedure was repeated: after each round of heating calculation, the melting depth was re-evaluated based on the threshold, and the conductivity profile was modified accordingly for the next iteration. All calculations ultimately converge to a stable melting structure and the heating profiles are shown in Figure~\ref{fig:melting}a,b.

\section{Effect of the post-perovskite phase on induction heating}\label{sec:ppv}

\begin{figure*}[ht!]
\centering
\includegraphics[width=\textwidth]{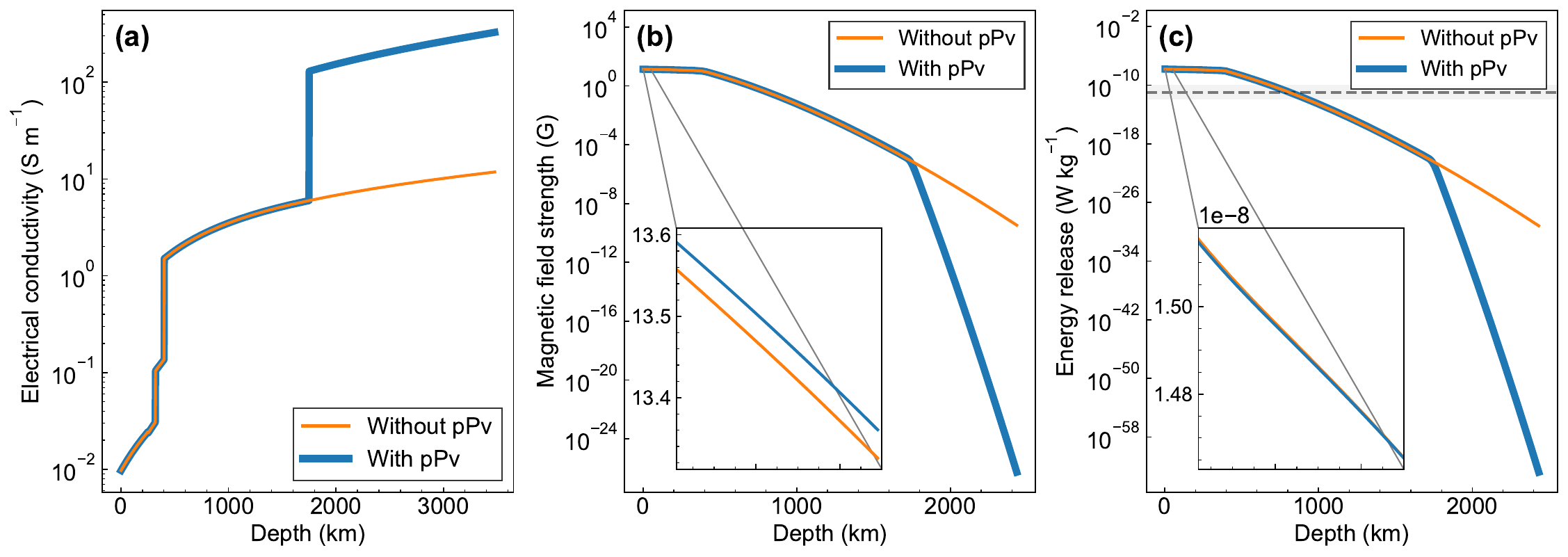}
\caption{The electrical conductivity (a), the modulus of the time-varying magnetic field (b), and the energy release per unit mass due to induction heating (c) as a function of depth inside GJ~486b with or without post-perovskite (pPv) phase in the conductivity profile.}\label{fig:ppv}
\end{figure*}

Bridgmanite undergoes a further phase transition to post-perovskite at $\sim$130 GPa \citep{Murakami2004a}, which is believed to be one of the dominant mineral phases in super-Earth mantles \citep[e.g.,][]{wagner_rocky_2012}. To date, the only available experimental study indicates that the electrical conductivity of (Mg$_{0.89}$Fe$_{0.11}$)SiO$_3$ post-perovskite is significantly higher than that of bridgmanite and shows negligible temperature dependence ($\sim$140 S\;m$^{-1}$ at 143 GPa) \citep{ohta_electrical_2008}. Although the data are insufficient to construct a robust conductivity model as a function of temperature, pressure, and composition, we adopt it as an anchor for a simplified estimate of mantle conductivity at pressures above 130 GPa. Specifically, we assume that the conductivity of post-perovskite is independent of temperature, increases linearly with pressure at a rate of 1 S\;m$^{-1}$ per GPa, and scales linearly with Fe content due to small polaron conduction being the dominant conduction mechanism \citep{ohta_electrical_2008}. Our calculation results suggest that the presence or absence of post-perovskite, as well as the specific choice of its conductivity, has almost no effect on the magnetic field strength and induction heating efficiency in the mantle regions below the phase transition pressure (Figure~\ref{fig:ppv}). Due to the fact that super-Earths experiencing significant induction heating are close-in exoplanets that are subject to high-frequency external magnetic fields, the skin depth of magnetic fields is usually small, and induction heating becomes negligible at the depths where post-perovskite is stable (Figure~\ref{fig:ppv}). Therefore, the potential increase in conductivity associated with the high-pressure phases in the deep mantles of super-Earths is unlikely to significantly impact their internal heat budget or thermal evolution via induction heating.

\section{Physical explanation of the induction heating profile}\label{sec:app-heating}

\begin{figure*}[ht!]
\centering
\includegraphics[width=0.8\textwidth]{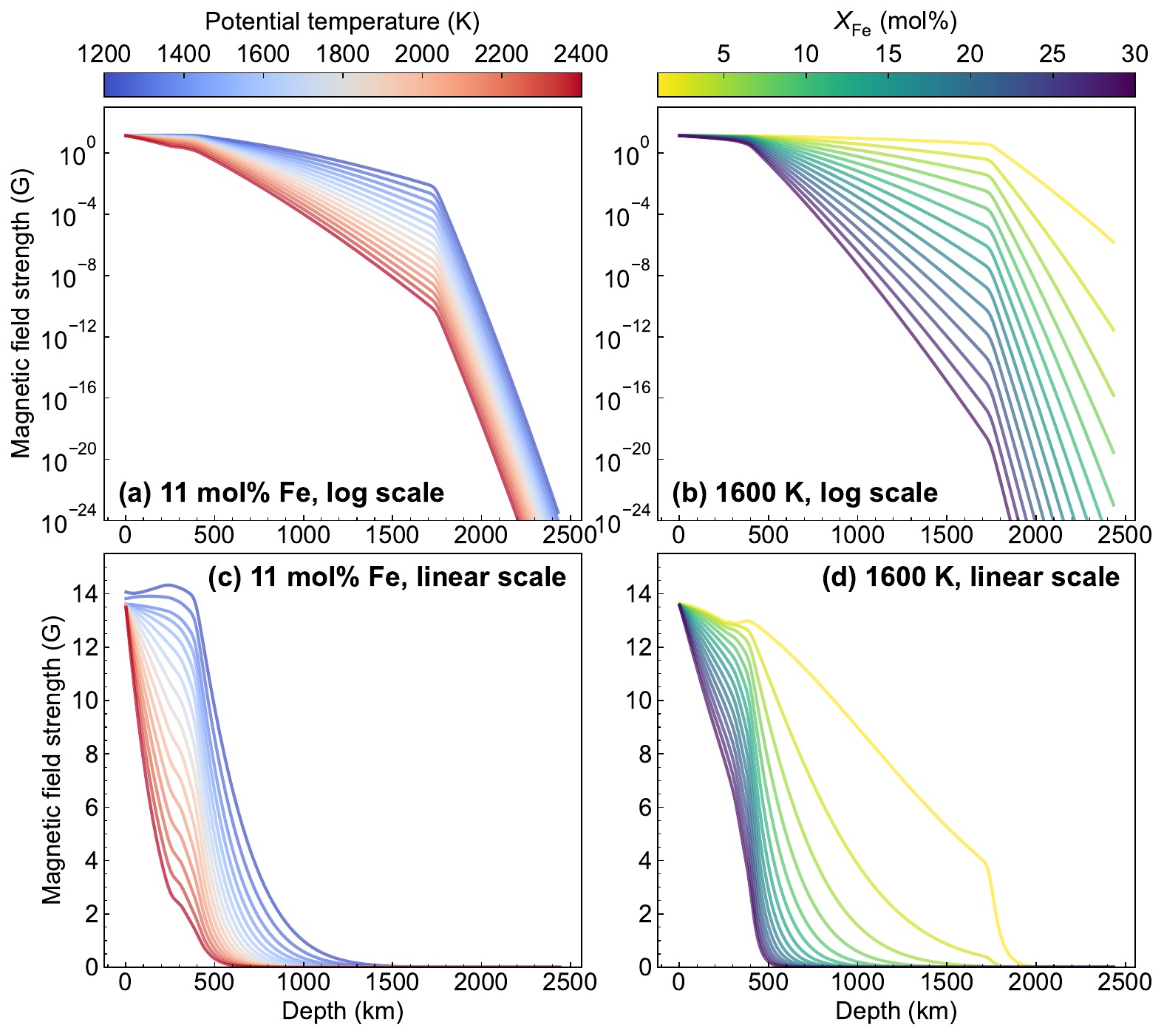}
\caption{The modulus of the time-varying magnetic field as a function of depth inside GJ~486b considering different potential temperatures (a, c) and Fe fractions in mantle silicates (b, d) in both log scale (a, b) and linear scale (c, d).}\label{fig:B}
\end{figure*}

Both temperature and pressure increase with depth within the planet, generally leading to a significant increase in electrical conductivity ($\sigma$), with abrupt transitions occurring at silicate phase boundaries. With a fixed magnetic field frequency, the skin depth ($\delta$) is determined solely by the conductivity of the medium, which in turn controls the efficiency of magnetic shielding---i.e., the exponential decay constant governing the attenuation of the magnetic field ($\bm{B}$) into the planet. As shown in Figure~\ref{fig:B}a,b, the slope of $\log(|\bm{B}|)$ vs. depth curve is determined by the conductivity of the stable mineral phase at the corresponding depth (see Figure~\ref{fig:sigma}).

\begin{figure*}[ht!]
\centering
\includegraphics[width=0.6\textwidth]{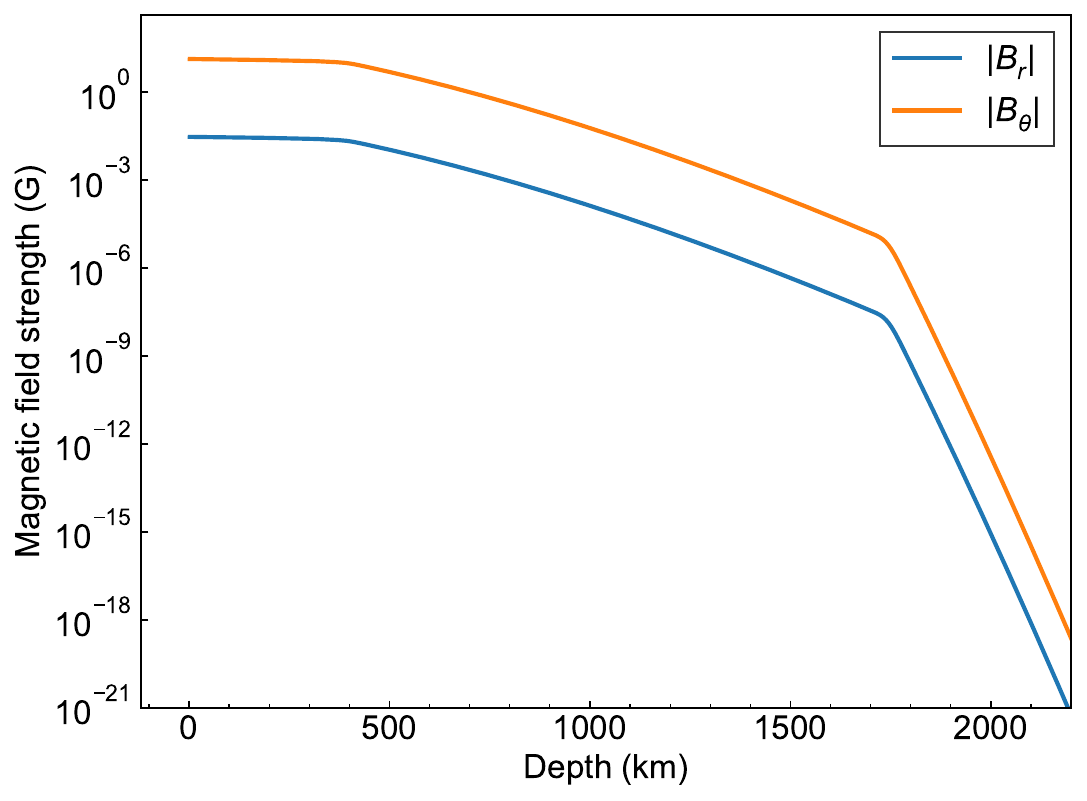}
\caption{The modulus of the $\theta$ and $\phi$ components of the time-varying magnetic field as a function of depth inside GJ~486b with a potential temperature of 1600 K and a mantle iron content of 11 mol\%.}\label{fig:Bdir}
\end{figure*}

\begin{figure*}[ht!]
\centering
\includegraphics[width=0.8\textwidth]{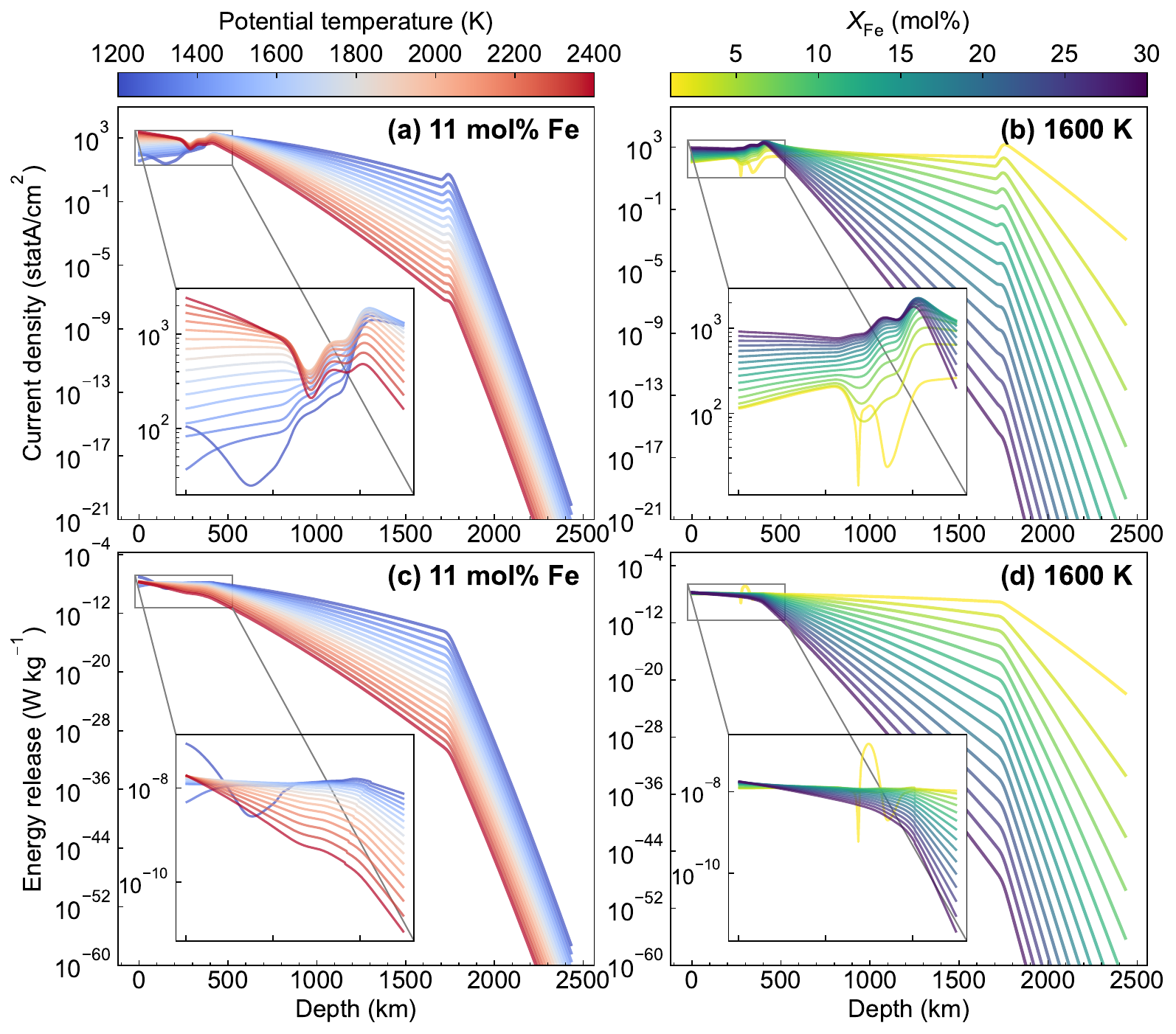}
\caption{The modulus of the induced current density (a, b) and energy release per unit mass due to induction heating (c,d) as a function of depth inside GJ~486b considering different potential temperatures (a, c) and Fe fractions in mantle silicates (b, d).}\label{fig:J}
\end{figure*}

Since the induced current density ($\bm{j}$) is proportional to the curl of the magnetic field ($\mu_0 \bm{j} = \nabla \times \bm{B}$), the spatial variation of the magnetic field dictates the distribution of current density. Given the dipolar nature of the magnetic field considered in this study, field lines bend primarily in the azimuthal direction as they enter the axisymmetric planet, and thus the internal magnetic field is overwhelmingly dominated by the angular component $B_\theta$ in spherical coordinates, with $B_\theta$ being approximately three orders of magnitude greater than $B_r$ (Figure~\ref{fig:Bdir}). Thus, the only nonzero component of the magnetic field curl, $(\nabla \times \bm{B})_\phi$, can be approximated by:
\begin{equation}\label{eq:curlB}
(\nabla \times \bm{B})_\phi = \frac{1}{r} \left[ \frac{\partial (rB_\theta)}{\partial r} - \frac{\partial B_r}{\partial \theta} \right] \approx \frac{\partial |\bm{B}|}{\partial r} + \frac{|\bm{B}|}{r}.
\end{equation}
In the near-surface region within the uppermost $\sim$1000 km, where induction heating is significant, $r$ is a large value close to the planetary radius ($\sim$8300 km), and the length scale of magnetic field variation is much smaller than the planetary radius. Therefore, the radial derivative term $\partial |\bm{B}|/\partial r$ typically dominates Equation \eqref{eq:curlB}. As a result, the profile of induced current density (Figure~\ref{fig:J}a,b) is largely controlled by the slope of the $|\bm{B}|$ vs. depth profile shown in Figure~\ref{fig:B}c,d.

The strong temperature dependence of upper mantle conductivity (Section~\ref{sec:sigma}) leads to a dramatic increase in the magnetic attenuation as $T_p$ increases (Figure~\ref{fig:B}c). At low $T_p$, olivine has low conductivity and thus provides weak magnetic shielding, leading to small values of both $\partial |\bm{B}|/\partial r$ and $|\bm{j}|$ in the upper mantle. When bridgmanite appears, the abrupt increase in $\sigma$ causes a sharp rise in $\partial |\bm{B}|/\partial r$ and a corresponding peak in $|\bm{j}|$. At greater depths, the decreasing magnitude of the magnetic field causes both $\partial |\bm{B}|/\partial r$ and $|\bm{j}|$ to decline continuously. In contrast, at high $T_p$, olivine becomes significantly more conductive, causing $|\bm{B}|$ to decay rapidly with a large radial gradient within the uppermost 500 km (Figure~\ref{fig:B}c), resulting in a high $|\bm{j}|$. Moreover, stronger magnetic shielding in the upper mantle at high $T_p$ reduces $|\bm{j}|$ in the deep mantle compared to low-$T_p$ conditions. As a result, the $|\bm{j}|$-depth curves for different $T_p$ values exhibit a clear crossover (Figure~\ref{fig:J}c).

Furthermore, the energy release due to Ohmic heating is given by
\begin{equation}\label{eq:energy}
    Q \propto |\bm{j}|^2/\sigma
\end{equation}
Therefore, the conductivity profile not only influences induction heating through its effect on current density but also directly modifies the energy release. For olivine in the upper mantle, the conductivity at $T_p = 2400$~K is three orders of magnitude higher than at $T_p = 1200$~K. This completely offsets the increase in $|\bm{j}|^2$ under high-$T_p$, high-$\sigma$ conditions, ultimately leading to a general decrease in induction heating with increasing $T_p$ in the upper mantle. As a result, the crossover in the $|\bm{j}|$ profiles does not persist in the energy release profiles (Figure~\ref{fig:J}a,c), and the general decrease in heat production with depth is also governed by the progressively increasing conductivity.

Similar effects of $X_{\mathrm{Fe}}$ on $\sigma$ and thus the profiles of $|\bm{B}|$, $|\bm{j}|$, and energy release can be observed in Figure~\ref{fig:B}b,d and Figure~\ref{fig:J}b, mirroring those of $T_p$. Because we considered only small polaron conduction for wadsleyite and ringwoodite, their conductivities are significantly lower than those of olivine (with ionic conductivity) and bridgmanite (with much higher small polaron conductivity than these phases) at low $ X_{\mathrm{Fe}}$ (Figure~\ref{fig:sigma}b). On the one hand, the abrupt decrease in conductivity causes fluctuations in the $|\bm{B}|$ vs. depth curve with $X_\mathrm{Fe} = 1$ mol\% at around 400 km depth (Figure~\ref{fig:B}d). Thus, two points where the derivative is zero occur in that region and cause two obvious minima in the corresponding curve in Figure~\ref{fig:J}b due to the $\partial |\bm{B}|/\partial r$ dependence of the current density (Equation~\eqref{eq:curlB}). On the other hand, the extremely low conductivity of the two phases directly enhances the Ohmic heating (Equation~\eqref{eq:energy}). Therefore, we find pronounced spikes in the current density and energy release curves for the $X_\mathrm{Fe} = 1$ mol\% within the depth ranges where wadsleyite and ringwoodite are stable (Figure~\ref{fig:J}b,d and Figure~\ref{fig:heating}b). Likewise, olivine exhibits extremely low conductivity at low temperatures due to its high activation energy, which leads to the strong fluctuations in the upper-mantle induction heating profiles with the lowest temperatures (Figure~\ref{fig:heating}a). At the planetary surface, $|\bm{B}|$ can be treated as constant and thus the exponential decay of $|\bm{B}|$ gives $\mathrm{d}|\bm{B}|/\mathrm{d}r \propto 1/\delta \propto \sqrt{\sigma}$. From Equation~\eqref{eq:curlB}, the current density scales as $|\bm{j}| \propto (\nabla \times \bm{B})_\phi \approx \mathrm{d}|\bm{B}|/\mathrm{d}r \propto \sqrt{\sigma}$, as long as the first term in Equation~\eqref{eq:curlB} is dominant (i.e., large enough near-surface slope in Figure~\ref{fig:B}cd). Substituting this relation into Equation~\eqref{eq:energy} yields $Q \propto |\bm{j}|^2 / \sigma \propto \sigma / \sigma$, indicating that, the energy release is usually almost independent of $\sigma$ at the planetary surface, which is clearly observed in Figure~\ref{fig:heating} and likely to hold for any planet.

Different from this study, previous models on induction heating in exoplanets have reported that the peak of the energy release occurs in the upper mantle, rather than exactly at the surface (Figure~\ref{fig:correct}) \citep[e.g.,][]{kislyakova_magma_2017, kislyakova_effective_2018, kislyakova_electromagnetic_2020}. This discrepancy arises from improvements in our induction heating model. Earlier calculations of Ohmic heating employed the relation $\nabla \times \bm{B} = \nabla^2 \bm{A}$ to derive the current density from the magnetic vector potential $\bm{A}$ \citep{kislyakova_magma_2017}. However, we found that this approach introduces large cancellation errors inherent in the analytical form of $\bm{A}$, leading to inaccurate estimates of current density and energy release near the surface. By adopting an updated model, we overcame these numerical inaccuracies and obtained more reliable results.

\section{Self-consistent temperature profiles}\label{sec:tempsc}
Due to the low thermal diffusivity of mantle rocks ($\kappa \sim 10^{-6}\;\mathrm{m^2\;s^{-1}}$), thermal conduction is inefficient at removing heat over lithospheric-to-mantle scales. For example, thermal homogenization at a length scale of 200 km would require more than 1.2 Gyr ($l = \sqrt{\kappa t}$). Therefore, sufficiently strong induction heating will lead to heat accumulation in the mantle, causing temperature increases and potentially partial melting. 

In Section~\ref{sec:molten} we have already explored, in a self-consistent manner, the thickness of partially molten layers and the corresponding induction heating efficiency. However, the modification of the mantle temperature profile itself by induction heating has not yet been explicitly calculated. In order to investigate the feedback of induction heating on the mantle temperature, even though conductive heat flow can be neglected, an additional outflow term must be taken into account: in a vigorously convecting mantle, heat generated by induction heating is carried away by moving material. The heat accumulation time $\Delta t$ in a convecting parcel of characteristic size $\Delta r$ is controlled by the convection turnover timescale, which allows us to estimate the net temperature increase of the parcel from the balance of heating and heat capacity:
\begin{equation}\label{eq:heating}
    \Delta T = \frac{E \Delta t}{C_p} = \frac{E \Delta r}{C_p v},
\end{equation}
where $E$ is the depth-dependent induction heating energy release per unit mass (in $\mathrm{W\;kg^{-1}}$), $C_p$ is the specific heat capacity of mantle rocks ($\sim 1000\;\mathrm{J\;kg^{-1}\;K^{-1}}$), and $v$ is the convection velocity. For a rough estimate, we adopt the terrestrial mantle velocity of $v \approx 5\;\mathrm{cm\;yr^{-1}}$ \citep{holzapfel_fe-mg_2005}. 

In constructing planetary interior profiles \citep{boujibar_super-earth_2020, zheng_cation_2025}, the adiabatic temperature gradient is expressed as
\begin{equation}
    \frac{dT}{dr} = - \frac{\rho g \gamma T}{K_s},
\end{equation}
where $\rho$ is density, $g$ is gravity acceleration, $\gamma$ is the Grüneisen parameter, and $K_s$ is the adiabatic bulk modulus. When accounting for additional heating by induction estimated in Equation \eqref{eq:heating}, the temperature gradient is modified to
\begin{equation}
    \frac{dT}{dr} = - \frac{\rho g \gamma T}{K_s} + \frac{E}{C_p v}.
\end{equation}
Thus, after computing an induction heating from an initial adiabatic temperature profile, we can reconstruct the temperature profile with the above expression and iterate until convergence is achieved. To avoid oscillatory behavior between higher and lower temperatures and accelerate convergence, the input profile at each iteration is taken as the average of the two previous iterations (a damping approach).

\begin{figure*}[ht!]
\centering
\includegraphics[width=0.8\textwidth]{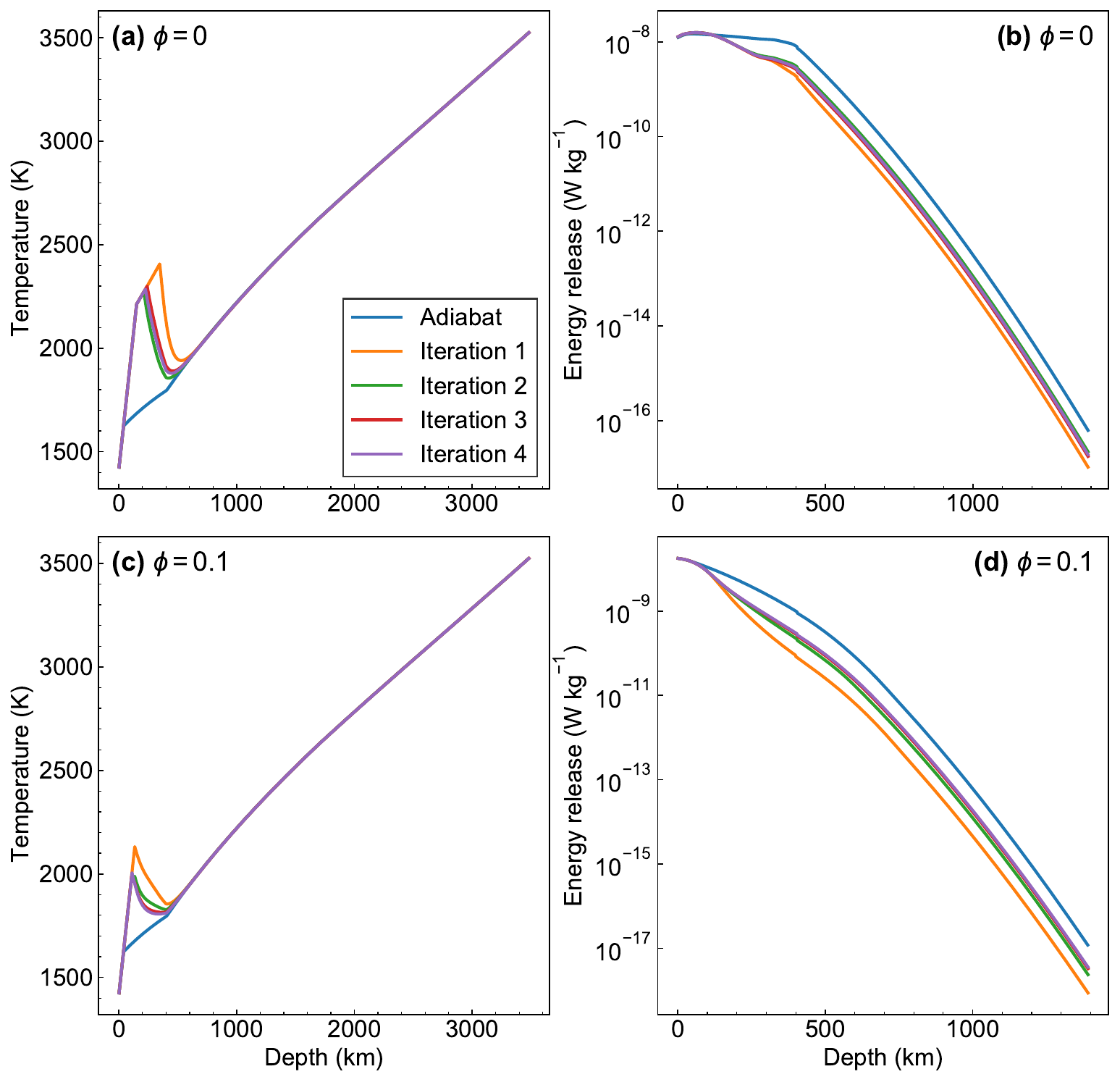}
\caption{The temperature profile (a,c) and energy release per unit mass due to induction heating (b,d) in GJ~486b obtained from different iterations of self-consistent calculation. The potential temperature and mantle iron content are 1600 K and 11 mol\%, respectively. (a,b) Fully solid mantle with melt fraction ($\phi = 0$). The total energy releases of the first and last iteration are 18.5~W\;m$^{-2}$ and 12.6~W\;m$^{-2}$, respectively. (c,d) partially molten mantle with melt fraction ($\phi = 0.1$). The total energy releases of the first and last iteration are 9.5~W\;m$^{-2}$ and 6.8~W\;m$^{-2}$, respectively. All temperatures are truncated at the mantle solidus \citep{hirschmann_mantle_2000, hirschmann_dehydration_2009, sahu_unveiling_2025}.}\label{fig:sc}
\end{figure*}

We applied this self-consistent scheme to the only planet in this study with significant induction heating, GJ~486b, adopting an Earth-like potential temperature of 1600 K and a mantle iron content of $X_\mathrm{Fe} = 0.11$ (11 mol\%). Tests were performed for melt fractions of 0 and 0.1. In both cases, convergence was reached within four iterations. Figure~\ref{fig:sc} shows the evolution of the temperature and induction heating profiles with iteration. The results suggest that the self-consistent induction heating profiles differ little from the one-shot calculations using an adiabatic profile: because of the higher near-surface temperatures, the total heating power is reduced by about one-third, but the overall heating characteristics remain unchanged. Therefore, incorporating a self-consistent temperature profile does not alter our conclusions.

It is important to note that once partial melting occurs near the surface, additional heat transport mechanisms become active, including melt percolation, volcanic heat loss, and thermal radiation. The percolation-induced heat flux can be over 500~W\;m$^{-2}$ at a melt fraction of 0.3 (in a Mars-sized body, which has a much smaller gravitational acceleration than super-Earths) \citep{zhang_two-phase_2021} and completely consume the energy released by subsequent induction heating, limiting the further temperature increase of the upper mantle after reaching the solidus temperature. For simplicity, we truncate the calculated temperature at the mantle solidus \citep{hirschmann_mantle_2000, hirschmann_dehydration_2009, sahu_unveiling_2025} in the near-surface partially molten region. The actual thermal structure in this region will be regulated by the interplay and feedback of multiple processes, and represents an important direction for future exploration.

\begin{figure*}[ht!]
\centering
\includegraphics[width=0.6\textwidth]{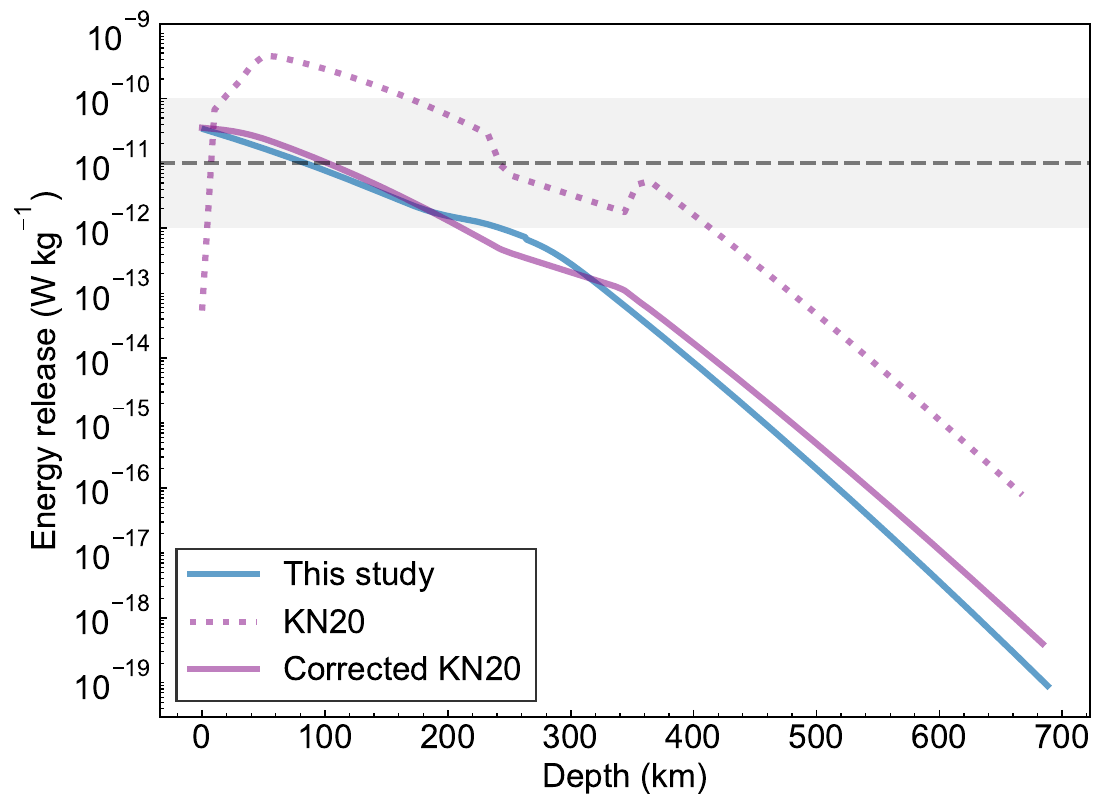}
\caption{Energy release per unit mass due to induction heating as a function of depth inside HD~3167b with a surface temperature of 2000 K (the potential temperature is 2400 K). KN20 represents the previously published numerically inaccurate results \citep{kislyakova_electromagnetic_2020}. Corrected KN20 represents the results obtained using the original conductivity profile of \citet{kislyakova_electromagnetic_2020} but with updated induction heating model.}\label{fig:correct}
\end{figure*}

\begin{figure*}[ht!]
\centering
\includegraphics[width=0.8\textwidth]{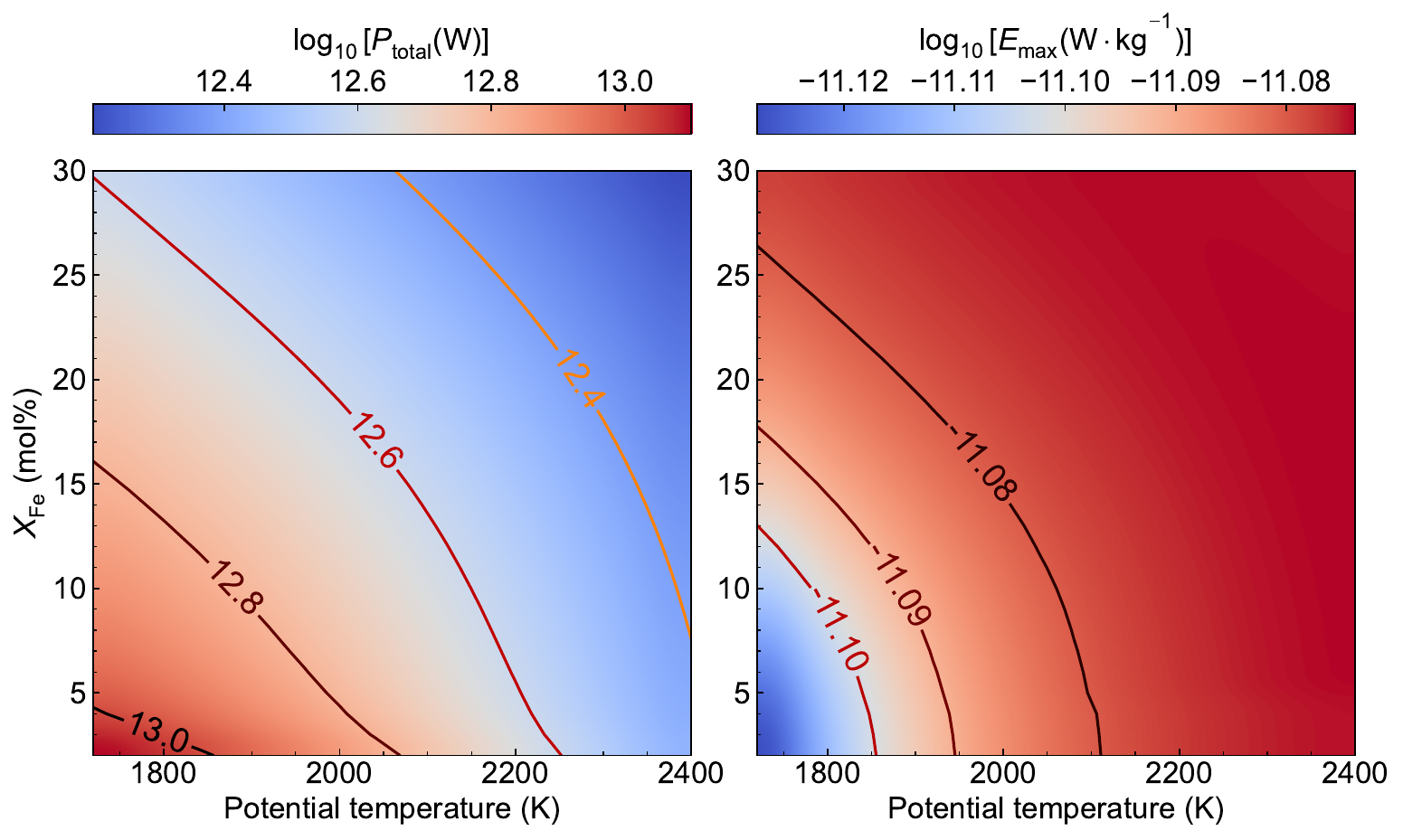}
\caption{Color maps and contour plots of induction heating efficiency as a function of potential temperature and mantle Fe contents ($X_\mathrm{Fe}$) in HD~3167b. (a) Total induction heating power inside the planet; (b) Maximum depth-dependent
energy release per unit mas}\label{fig:5G}
\end{figure*}

\begin{figure*}[ht!]
\centering
\includegraphics[width=0.6\textwidth]{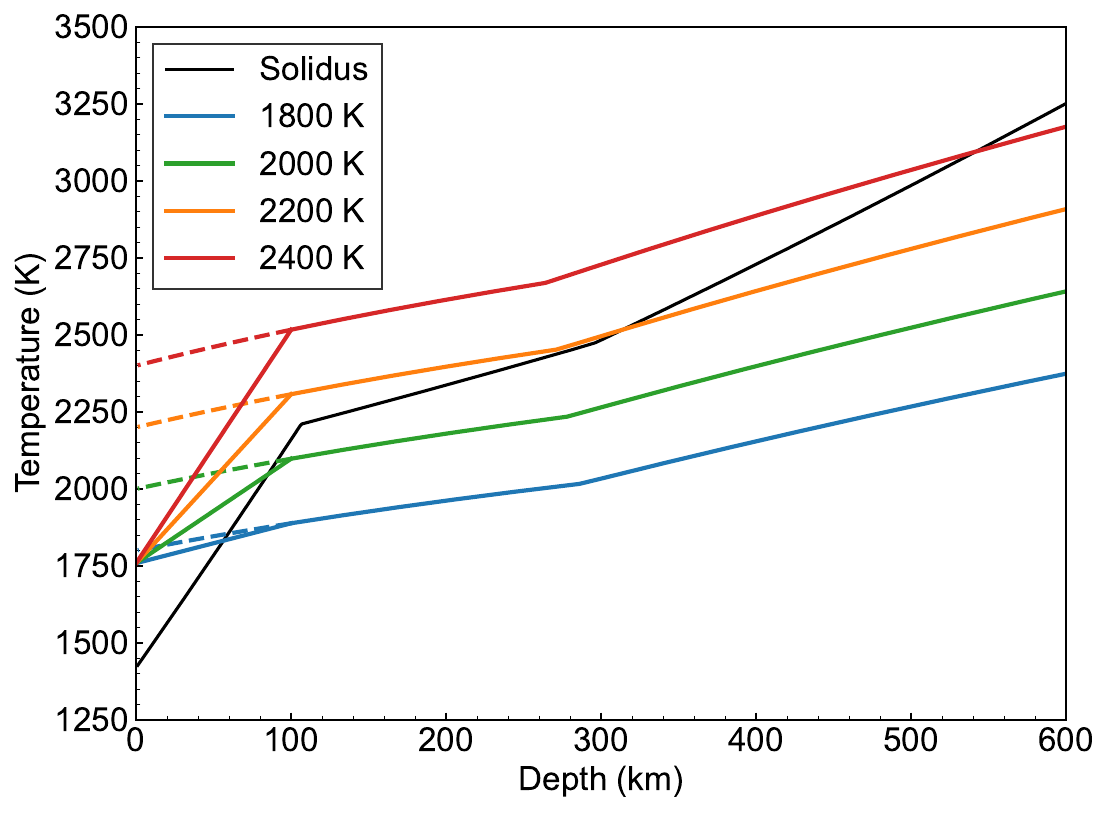}
\caption{The temperature profile of HD~3167b obtained with different potential temperatures, in comparison with the mantle solidus \citep{hirschmann_mantle_2000, hirschmann_dehydration_2009, sahu_unveiling_2025}.}\label{fig:solidus}
\end{figure*}


\clearpage

\bibliography{references}{}
\bibliographystyle{aasjournal}


\end{document}